\documentclass[prl,aps,twocolumn,showpacs,superscriptaddress,floatfix,amsmath,amssymb,nofootinbib]{revtex4}
\usepackage{amsfonts}
\usepackage{graphicx} %con este paquete podemos meter graficos
\newcommand{\beq}{\begin{equation}}
\newcommand{\eeq}{\end{equation}}

\newcommand{\beqa}{\begin{eqnarray}}
\newcommand{\eeqa}{\end{eqnarray}}
\newcommand{\beqas}{\begin{eqnarray*}}
\newcommand{\eeqas}{\end{eqnarray*}}
\def\ra{\rangle}
\def\la{\langle}
\begin{document} %aqui empiezas el documento
\title{Fast transport of Bose-Einstein condensates}
\author{E. Torrontegui}
\affiliation{Departamento de Qu\'{\i}mica F\'{\i}sica, Universidad del Pa\'{\i}s Vasco - Euskal Herriko Unibertsitatea, 
Apdo. 644, Bilbao, Spain}

\author{Xi Chen}
\affiliation{Departamento de Qu\'{\i}mica F\'{\i}sica, Universidad del Pa\'{\i}s Vasco - Euskal Herriko Unibertsitatea, 
Apdo. 644, Bilbao, Spain}
\affiliation{Department of Physics, Shangai University, 200444 Shanghai, P. R. China}

\author{M. Modugno}
\affiliation{Departamento de F\'{\i}sica Te\'orica e Historia de la Ciencia, Universidad del Pa\'{\i}s Vasco - Euskal Herriko Unibertsitatea, 
Apdo. 644, Bilbao, Spain}
\affiliation{IKERBASQUE, Basque Foundation for Science, Alameda Urquijo 36, 48011 Bilbao, Spain}

\author{S. Schmidt}
\affiliation{Institut f\"ur Theoretische Physik, Leibniz
Universit\"{a}t Hannover, Appelstra\ss e 2, 30167 Hannover,
Germany}

\author{A. Ruschhaupt}
\affiliation{Institut f\"ur Theoretische Physik, Leibniz
Universit\"{a}t Hannover, Appelstra\ss e 2, 30167 Hannover,
Germany}

\author{J. G. Muga}
\affiliation{Departamento de Qu\'{\i}mica F\'{\i}sica, Universidad del Pa\'{\i}s Vasco - Euskal Herriko Unibertsitatea, 
Apdo. 644, Bilbao, Spain}
\begin{abstract}
We propose an inverse method to accelerate without final excitation the adiabatic
transport of a Bose Einstein condensate. The method, applicable to arbitrary
potential traps, is based on a partial extension of the Lewis-Riesenfeld
invariants, and 
%In contrast with trap expansions or compressions, for which the inverse
%  method applied to condensates is constrained to 2D traps, or to a
%  Thomas-Fermi regime, or it requires time-dependent Feshbach resonance
%  manipulation, 
provides transport protocols that satisfy exactly the no-excitation
conditions without constraints or approximations. This inverse method is complemented by optimizing the trap trajectory with respect to different physical criteria and by studying the effect of noise.

\end{abstract}  	
\maketitle
{\it Introduction.}---A
major goal of atomic physics is the comprehensive control of the
atomic quantum state for fundamental research and applications in interferometry, metrology, or information processing. 
The ability to manipulate Bose-Einstein condensates may be particularly
rewarding due, for example, to their potential in interferometric sensors, but
it is also challenging, as their low temperatures make them more fragile than
ordinary cold atoms.
A basic operation is the transport of the condensate to appropriate locations such as a ``science chamber'', or to launch or stop the atomic cloud. This transport has been performed with several techniques  based on adiabatic, slow motion to avoid excitations and losses \cite{Ketterle2002,H_Nature,Denschlag06,Xiong}. Long transport times, however, may be counterproductive since the condensate is more exposed to noise and decoherence, and also limit severely the repetition rates and signal to noise ratios. 
Fast, non-adiabatic but ``faithful'' transport of cold atoms, i.e., leading to
the desired final state, has also been investigated experimentally \cite{David}
and theoretically \cite{Calarco,Masuda,transport}. For the Schr\"odinger equation (SE) an inverse engineering method based on constructing Lewis-Riesenfeld invariants  and corresponding dynamical modes (solutions of the SE formed by invariant eigenvectors times a phase factor) provides a ``shortcut to adiabaticity'' \cite{ChPRL1,Nice}.
However the invariant concept, i.e., 
an operator satisfying $dI/dt\equiv
{\partial I(t)}/{\partial t} +[I(t),H(t)]/({i\hbar})=0$ with constant expectation values for arbitrary states that evolve with the Hamiltonian $H$, is not directly applicable to the non-linear Gross-Pitaevskii equation (GPE). In fact previous extensions of this inverse technique to expansions of condensates required special regimes or time-dependent Feshbach resonance control \cite{MCh,Nice2,Adolfo}. By contrast, we show in this letter that the transport of a condensate scales in the right way for applying a generalized inverse method, so that no approximations or limits are required to design fast processes without final excitation. Strictly speaking we do not construct invariants as in the linear theory but, as far as ground-state to ground-state transport is concerned, the formal structure of the dynamical transport modes of the linear case remains valid for the GPE. This will be illustrated with a numerical example and compared with a direct method. In the final part of the article we shall optimize the trap trajectory according to several criteria, since  
the invariant-based inverse engineering provides 
a family of possible transport solutions, an analyze the effect of noise.    

{\it General setting.}---Our
starting point is the GPE for potentials whose Schr\"odinger dynamics admit a quadratic invariant 
in momentum \cite{LL,transport,Lohe},
\beqa
&&i\hbar\frac{\partial\psi ({\mathbf q},t)}{\partial t}=\bigg[-\frac{\hbar^2}{2m}\nabla^2_{{\mathbf q}}
-{\bf F}(t)\cdot{\mathbf q}+
\frac{1}{2}m\omega^2(t)|{\bf q}|^2
\nonumber \\
&&+\frac{1}{\rho^2}U\bigg(\frac{{\bf q}-\boldsymbol{\alpha}}{\rho}\bigg)+g_{_D}|\psi({\mathbf q},t)|^2+f(t)\bigg]\psi({\mathbf q},t),
\label{GP}
\eeqa
where $f=f(t)$ is arbitrary, $\nabla^2_{\mathbf q}$ is the Laplacian in Cartesian coordinates for $D=1$, 2, or 3 dimensions, $U$ 
is an arbitrary potential function of the argument ${\boldsymbol \sigma}\equiv({\mathbf q} -\boldsymbol{\alpha})/\rho$ 
and $\omega (t)$, the force ${\mathbf F}(t)$, $\boldsymbol{\alpha}=\boldsymbol{\alpha} (t)$ and the scaling
function $\rho=\rho(t)$ satisfy
\beqa
\omega_0^2/\rho^3&=&\ddot\rho+\omega^2(t)\rho,
\label{Ermakov}
\\
{\mathbf F}(t)/m&=&\ddot{\boldsymbol{\alpha}}
+\omega^2(t)\boldsymbol{\alpha},
\label{oscilador} 
\eeqa
where $\omega_0$ is a constant and the dots represent time derivatives. The physical meaning of $\boldsymbol{\alpha}$
depends on the process type, see \cite{transport}, and will be clarified below. The wave function is normalized to one. 
Without the non-linear term, an arbitrary solution of Eq. (\ref{GP}) can be written as a linear combination of the eigenvectors 
$\psi_n$  of the dynamical invariant $I$ \cite{LR}, 
%
%\beqas
$\psi(\mathbf q,t)=\sum_n c_n e^{i\alpha_n(t)} \psi_n(\mathbf q,t)$,  %\label{lewis}
%\\
$I(t)\psi_n(\mathbf q,t)=\lambda_n\psi_n(\mathbf q,t)$,
%\label{inva}
%\eeqas
%
where the amplitudes $c_n$  and the eigenvalues $\lambda_n$ are constants.  
Here $\psi_n$ is normalized to one, but continuum, delta-normalized states are also possible.   
The Lewis-Riesenfeld phases $\alpha_n$ satisfy
%
%\beq
$\hbar\, \frac{d \alpha_n}{dt}=\left\la\psi_n\left|i\hbar\frac{\partial}{\partial t}
-H\right|\psi_n\right\ra$ \cite{LR,DL}.
%\eeq
A single mode solution of the SE ($g_{_D}=0$)
takes the form
\beqa
&&e^{i\alpha_n}\psi_n({\mathbf q},t)=\rho^{-D/2}
e^{\frac{im}{\hbar\rho}[\dot\rho|{\mathbf q}|^2/2+(\dot{\boldsymbol{\alpha}}\rho-\boldsymbol{\alpha}\dot\rho)\cdot{\mathbf q}]}
\nonumber\\
&&\times e^{-\frac{i}{\hbar}\int_{0}^{t}\!\! dt'\Big\{\frac{m\left[|\dot{\boldsymbol{\alpha}}\rho-\boldsymbol{\alpha}\dot\rho|^2-\omega_0^2|\boldsymbol{\alpha}|^2/\rho^2\right]}{2\rho^2}+f\Big\}}\phi_n(\boldsymbol{\sigma},\tau),
\label{psin}
\eeqa
where we have introduced a scaled time $\tau=\int_{0}^{t} dt'\rho^{-2}$,
%$D=1,2,3$ depending on the dimension,
and $\phi_n$ satisfies a Schr\"odinger
equation with a time-independent Hamiltonian.
%
%\beqas
%&&i\hbar\frac{\partial\phi_n ({\boldsymbol{\sigma}},\tau)}{\partial\tau}
%\nonumber
%\\
%&=&\bigg[-\frac{\hbar^2}{2m}\nabla^2_{\boldsymbol{\sigma}}+
%\frac{1}{2}m\omega_{0}^{2}|\boldsymbol{\sigma}|^2+U(\boldsymbol{\sigma})
%\bigg]
%\phi_n({\boldsymbol{\sigma}},\tau),
%\label{stats} 
%\eeqas
%
%where we have eliminated the irrelevant time-dependent global term
%$-m\omega_{0}^{2}|\boldsymbol\alpha|^2/(2\rho^2)$.

These results can be generalized partially   
for the GPE, and the extent of the generalization depends on the process type.
Inserting Eq. (\ref{psin}) as an ansatz for a time dependent solution of the GPE, $\phi_n$ must satisfy 
\beqa
i\hbar\frac{\partial\phi ({\boldsymbol{\sigma}},\tau)}{\partial\tau}&=&\bigg[-\frac{\hbar^2}{2m}\nabla^2_{\boldsymbol{\sigma}}+
\frac{1}{2}m\omega_{0}^{2}|\boldsymbol{\sigma}|^2+U(\boldsymbol{\sigma})
\nonumber \\
&+&\rho^{2-D}g_{_D}|\phi ({\boldsymbol{\sigma}},\tau)|^2\bigg]\phi ({\boldsymbol{\sigma}},\tau).
\label{stat} 
\eeqa
Unlike the linear theory, we cannot construct the general solution 
by linear superposition, so we restrict the treatment to a single mode, for
example the ground state, and therefore we drop the $n$ subindex in Eq. (\ref{stat}) and hereafter. 

Equation (\ref{stat}) is very general and applicable to compressions, expansions, or transport for harmonic or anharmonic potentials. 
It is most useful when $\rho^{2-D} g_{_D}$ does not depend on time, since the physical solution of the time-dependent 
problem is then mapped, via Eq. (\ref{psin}), to the solution
of a much simpler stationary equation. This happens in several physically relevant cases, in particular for expansions of BECs when 
$D=2$, or by tuning $g_{_D}$ as a time-dependent coupling to cancel 
the time dependence of $\rho^{2-D}$ \cite{MCh}. 
Different time scalings combined with a Thomas-Fermi approximation also lead
to a stationary equation \cite{MCh}. 

{\it Transport Processes.}---Here
we are interested in the simple but very important case 
$\rho (t)=1$ $\forall t$, associated with rigid transport processes. Then $\tau=t$ and the coefficients of Eq. (\ref{stat}) are time independent. It is also useful to define $\phi(\boldsymbol\sigma,t)=e^{-i\mu t /\hbar}\chi(\boldsymbol\sigma)$,
where $\mu$ is the chemical potential and $\chi(\boldsymbol\sigma)$ satisfies the stationary GPE
\beqa
&&\bigg[-\frac{\hbar^2}{2m}\nabla^2_{{\boldsymbol\sigma}}+
\frac{1}{2}m\omega^{2}_{0}|{\boldsymbol\sigma}|^2+U(\boldsymbol\sigma)
+g_{_D}|\chi(\boldsymbol\sigma)|^2\bigg]\chi(\boldsymbol\sigma) \nonumber \\
&=&\mu\chi(\boldsymbol\sigma).
\label{GPstationary}
\eeqa
The physical solution of the time-dependent
GPE equation for a single transport mode is
\beq
\psi(\mathbf{q},t)=e^{\frac{i}{\hbar}\{-\mu t+m\boldsymbol{\dot{\alpha}}\cdot\mathbf q -\int_{0}^{t}\!dt'[\frac{m}{2}(|\boldsymbol{\dot{\alpha}}|^2-\omega_0^2|\boldsymbol{\alpha}|^2)+f]\}}
\chi(\boldsymbol\sigma), 
\label{trans}
\eeq
which is a fundamental result. Since this wave function is shape invariant the only possible excitations associated with such a mode are center of mass oscillations with constant mean field energy. 
%This does not depend on the dimension $D$. 
In the following we shall apply it to 1D transport and 
omit vector notation. 
%As we are interested in the
%transport process between an initial point and a final point,
%we restrict in the following only at the one-dimensional transport scenario
%between these two points an omit the vector notation. 

{\it Inverse engineering by harmonic transport.}---In
%\paragraph{Inverse engineering}
%
%
%
a 1D, horizontal, harmonic  transport of a condensate from $0$ to $d$ in a time $t_f$ with zero mean velocity at $t=0$ and $t_f$, $\alpha$ (now a scalar function) must 
be chosen to match the transport mode  (\ref{trans}) with the 
instantaneous eigenstates of the Hamiltonian, including the mean field term, at times $t=0$ and $t=t_f$. 

A concrete example will illustrate how this works. 
Consider a $^{87}$Rb BEC in the $F=2$, $m_F=2$
ground state constituted by $3000$ atoms \cite{H_Nature}. The transport of BECs aided by microchips can use a ``bucket chain'' \cite{H_PRL}, or a single harmonic and frequency-stable bucket \cite{Zimmermann05}. We assume here a single 
bucket with $\omega_0=2\pi\cdot 50$ Hz
moved from $q_0(0)=0$ at time $t=0$ to $q_0(t_f)=d=1.6$ mm at $t_f$. 
The time-dependent GPE is now 
\beqa
i\hbar\frac{\partial\psi (q,t)}{\partial t}&=&\bigg[-\frac{\hbar^2}{2m}\nabla^2_{q}+\frac{1}{2}m\omega_{0}^{2}(q-q_{0})^2
\nonumber\\
&+&g_{_{1}}|\psi (q,t)|^2\bigg]\psi(q,t),
\label{GP2} 
\eeqa
a particular case of Eq. (\ref{GP}) with 
$\omega(t)=\omega_0$, $U=0$, $F(t)=m\omega_{0}^{2}q_{0}(t)$, 
$f(t)=m\omega_0^2q_0^2(t)/2$ and
$\rho(t)=1$, 
%$\forall t$,
so Eq. (\ref{Ermakov}) does not play any role and
$\alpha=q_c$ has to satisfy
\beq
\ddot q_c+\omega_{0}^{2}(q_c-q_0)=0,  
\label{osc}
\eeq
the equation for a classical trajectory $q_c(t)$ in a moving harmonic potential.  (Note that for an abrupt shift of the trap one recovers the scaling for dipole oscillations 
\cite{Dalfovo}.) 
If we impose at $t=0$ the initial conditions
\beq
q_c(0)=\dot q_c(0)=\ddot q_c(0)=0,
\label{inicial}
\eeq
the transport mode (\ref{trans}) becomes equal to the instantaneous eigenstates of Eq. (\ref{GP2}) at $t=0$.    
%
%
%
%%%%%%%%%%%%%%%%%%%%%%%%%%%%%%%%%%%%%%%%%%%%%%%%%%%%%%%%%%%%%%%%%%%%
\begin{figure}[t]
\begin{center}
\includegraphics[width=0.8\linewidth,height=3.5cm]{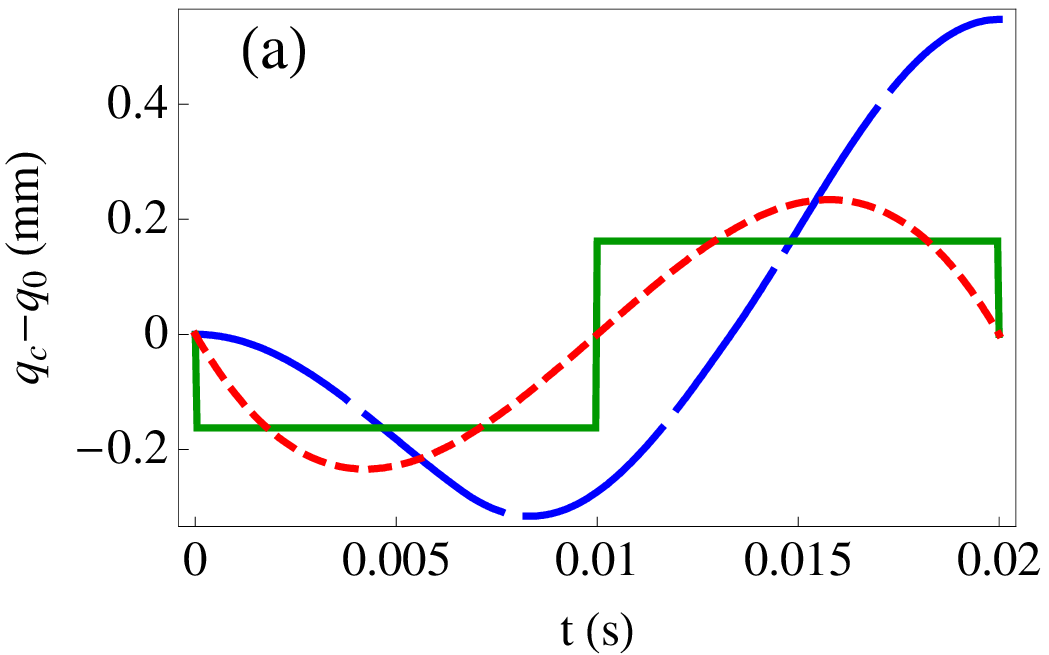}

\includegraphics[width=0.8\linewidth,height=3.5cm]{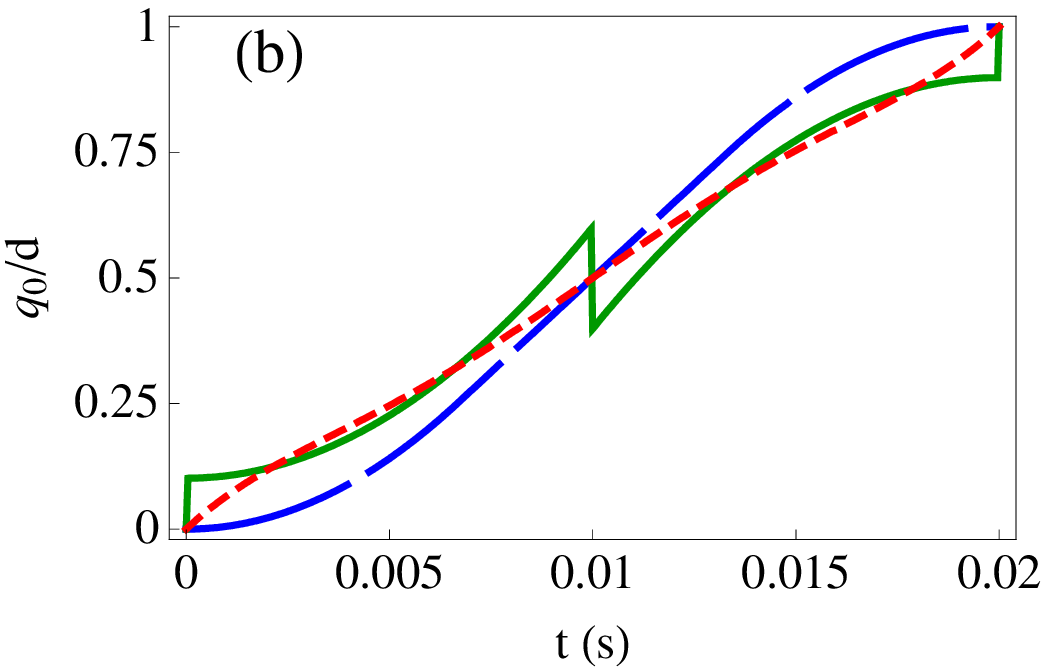}

\caption{(Color online) (a) Displacement $q_c-q_0$ versus time. 
Blue long dashed line: direct method,
red short dashed line: inverse method (polynomial). Solid green line: inverse method + OCT. 
(b) Trap trajectories. Parameter values: $d=1.6$ mm, $t_f=20$ ms,
$\delta\simeq 0.162$ mm, $\omega_0=2\pi\cdot 50$ Hz.}
\label{fig1}
\end{center}
\end{figure}
%%%%%%%%%%%%%%%%%%%%%%%%%%%%%%%%%%%%%%%%%%%%%%%%%%%%%%%%%%%%%%%%%%%%%%
% 
%
%
To solve Eq. (\ref{GP2}) we proceed in two different ways, using  direct or
inverse approaches.

In the direct
approach we fix first the evolution of the center of the trap $q_0(t)$.  
In \cite{H_Nature}, for example, see Fig. 5 there, $\dot{q}_0(t)$  
is increased linearly during a quarter of the transported distance $d/4$, then kept constant for 
$d/2$, and finally ramped back to zero during the last quarter,
\beqas
q_{0}(t)=\left\{
\begin{array}{ll}
{\frac{v_{m}^{2}t^2}{d}},& 0<t<\frac{d}{2v_m}
\\
v_mt-\frac{d}{4},&
\frac{d}{2v_m}<t<\frac{d}{v_m}
\\
\frac{v_m}{2(d/v_m-t_f)}(t-t_f)^2+d,&
\frac{d}{v_m}<t<t_f
\end{array}
\right.,
\eeqas
where $v_m=3d/(2t_f)$ is the maximum trap velocity during the transport 
compatible with $q_0(t_f)=d$ in this scheme.
Solving Eq. (\ref{osc}) for the previous $q_0(t)$ with initial conditions $q_c(0)=\dot q_c(0)=0$, and imposing 
continuity on $q_c(t)$ and $\dot q_c(t)$ we find 
\beqa
q_c(t_f)\!-\!q_0(t_f)\!&=&\!{9d}(1-2\cos\phi)(\sin^2 \phi)/(\omega_{0}^{2}t_{f}^{2}),
\nonumber \\
\dot q_c(t_f)\!-\!\dot q_0(t_f)\!&=&\!\frac{9d}{2\omega_{0}t_{f}^{2}}(\sin\phi
+\sin 2\phi -\sin 3\phi), 
\label{qH}
\eeqa
where $\phi=\omega_0 t_f/3$. 
The final state of the transported BEC
is given by Eqs. (\ref{trans}) and (\ref{qH}). In general
some excitation is produced, except for the discrete set of final times $t_{f,N}=3(2N+1)\pi/\omega_0$, $N=0,1,2, ...$, for which 
\beq
q_c(t_f)=d,\quad \dot q_c(t_f)=\ddot q_c(t_f)=0,
\label{final}
\eeq
and the transported state matches the eigenstate of the final Hamiltonian. 
The classically moving center of mass and the trap center stop at $d$, $\dot q_c(t_f)=0$, $\dot q_0(t_f)=0$, 
with zero (classical) energy $m\dot q_c(t_f)^2/2+m\omega_{0}^{2}[d-q_c(t_f)]^2/2=0$.
Using this direct approach, the minimum final time which does not produce
excitation is $t_{f,0}=3\pi/\omega_0$ ($N=0$). 
In our example, $t_{f,0}=30$ ms. 
%This result goes beyond a simple slow adiabatic approach since 
For such short times the transport is not adiabatic. 

Thanks to the structure of the solution 
(\ref{trans}), we may apply a generalized inverse engineering method 
similar to the one for the linear case \cite{MCh, ChPRL1, transport}. 
The idea is to design $q_c(t)$ first and deduce the transport
protocol from it.
We impose the conditions (\ref{inicial}) and (\ref{final}) at $t=0$ and $t_f$, and interpolate $q_c$ with a function, e.g. a polynomial with enough parameters to  satisfy all these conditions. Then $q_0(t)$ is calculated 
via Eq. (\ref{osc}). An example is shown in Fig. 1
%In Fig. \ref{fig1} we plot the mean potential energy $\langle V \rangle$, including the mean-field interaction, 
%of the transported state versus time. The initial and final state is the ground state of Eq. (\ref{GP2}). Eq. (\ref{GPstationary}) is solved by a Fast-Fourier-Transform  propagation in imaginary time.  
where we have chosen $t_f=20$ ms $<t_{f,0}$. 
By construction no final excitation is produced, and the final fidelity (overlap between the transported
state and the ground state at $t_f$) is one. Contrast this to  
the direct approach which, for $t_f=20$ ms, produces more transient excitation and a final excited state with nearly zero fidelity.    

In principle there is no lower limit to $t_f$ with the inverse method, but in practice 
there are some limitations \cite{transport}. Smaller values of $t_f$ increase the distance from the condensate to the trap center, see Eq. (\ref{qH}), and the effect of anharmonicity. There could be also geometrical constraints: for short $t_f$, $q_0(t)$ 
could exceed the interval [$0,d$]. For the polynomial ansatz  this happens \cite{transport} at $t_f=2.505/\omega_0$, $t_f\approx 8$ ms
for the parameters of the example. Optical Control Theory (OCT) combined 
with the inverse method, see below, provides a way to design 
trajectories taking these restrictions into account.  
 
{\it Anharmonic Transport.}---The
inverse method can also be applied to anharmonic transport by means of a compensating force \cite{transport}. To this aim, we consider 
a generic potential $U(q-q_0)$ and set 
$\alpha(t)=q_0(t)$, $\omega(t)=\omega_0=0$, $f=0$, and $F=m\ddot q_{0}$ in Eq. (\ref{GP}), so the GPE becomes 
\beqas
&&i\hbar\frac{\partial\psi (q,t)}{\partial t}=\bigg[-\frac{\hbar^2}{2m}\nabla^2_{q}-
m\ddot q_0q+U(q-q_0) \nonumber \\
&&+g_{_1}|\psi(q,t)|^2\bigg]\psi(q,t),
%\label{GP3}
\eeqas
and the auxiliary equations (\ref{Ermakov}) and (\ref{oscilador}) are satisfied trivially.
Here we impose
%
%\beqas
$q_0(0)=0, \quad\dot q_{0}(0)=0$,
%\nonumber \\
$q_0(t_f)=d, \quad\dot q_{0}(t_f)=0$.
%\eeqas
%
We may optionally impose also $\ddot q_0=0$ at $t=0$ and $t_f$. 
The function that must be interpolated is now $q_0(t)$, and again we 
consider a polynomial.  
For an arbitrary trap and $t_f=20$ ms, the maximal 
compensating acceleration would be $23.1$ m/s$^2$.
% Note that the exact potential $U$ could be unknown. 
%
%
%

{\it Optimal control theory.}---Given 
the freedom left by the inverse method it is 
natural to combine it with OCT and design the
trajectory according to relevant physical criteria \cite{stef}. 
For harmonic transport, 
we have imposed the boundary conditions (\ref{inicial}) and (\ref{final}) at $t=0$ and $t_f$, but $q_0(t)$, and the polynomial ansatz for $q_c(t)$
are quite arbitrary. As an example of the possibilities of OCT 
suppose that we wish to limit the 
deviation of the condensate from the trap center 
according to   
$-\delta\leq q_c-q_0\leq \delta$,
$\delta>0$ and find the minimal time $t_f$. 
The transport process given by Eqs. (\ref{trans}), (\ref{inicial}) and (\ref{final}) can be rewritten as a minimum-time optimal control problem
defining the state variables $x_1(t)$ and $x_2(t)$ and the control $u(t)$,
\beq
\label{def}
x_1=\alpha, \quad x_2={\dot\alpha}, \quad u(t)=\alpha-q_0.
\eeq
Equation (\ref{osc}) is transformed into a system of equations, 
\beq
\label{oct}
\dot x_1=x_2, \quad \dot x_2+\omega_0^2u=0.
\eeq
The OCT problem is to find $-\delta\leq u(t)\leq \delta$ with $u(0)=u(t_f)=0$,  $\{x_1(0),x_2(0)\}=\{0,0\}$, and $\{x_1(t_f),x_2(t_f)\}=\{d,0\}$ in the minimum final time $t_f$. The optimal control Hamiltonian \cite{LSP} is
%
%\beqas
$H_c=p_1 x_2-p_2 \omega_0^2u$,
%\eeqas
%
where $p_1$ and $p_2$ are conjugate variables. The Pontryagin maximality principle \cite{LSP} tells us
that for $u(t)$, $\mathbf x(t)$ to be time-optimal, it is necessary that there exists a nonzero, continuous vector
$\mathbf p(t)$ such that $\mathbf{\dot x}=\partial H_c/\partial\mathbf p$, $\mathbf{\dot p}=-\partial H_c/\partial\mathbf x$
at any instant, the value of the control maximizes $H_c$, and $H_c(\mathbf p(t),\mathbf x(t), \mathbf u(t))=c\geq 0$, with $c$
a constant. 
The solution is of bang-bang type \cite{Salamon09}, 
\beqas
u(t)=\left\{
\begin{array}{ll}
0,& t=0
\\
-\delta,& 0<t<t_1
\\
\delta,& t_1<t<t_f
\\
0, & t=t_f
\end{array}
\right.,
\eeqas
where the initial and final discontinuities are chosen to satisfy the boundary conditions. 
Solving the system (\ref{oct}) and imposing
continuity on $x_1$ and $x_2$ one finds for the switching and final times
%
%\beq*
$t_1={t_f}/{2}$, $t_f={2}({d}/{\delta})^{1/2}/{\omega_0}$.
%\eeq*
%
The trap trajectory is deduced from Eq. (\ref{def}), 
\beqas
q_0(t)=\left\{
\begin{array}{ll}
0,&t=0
\\
({1+\omega_{0}^{2}t^2}/{2})\delta,& 0<t<t_1
\\
-\big[{\omega_{0}^2}(t-t_f)^2/2+1\big]\delta+d,& t_1<t<t_f
\\
d,&t=t_f
\end{array}
\right..
\eeqas
%
%
%
%%%%%%%%%%%%%%%%%%%%%%%%%%%%%%%%%%%%%%%%%%%%%%%%%%%%%%%%%%%%%%%%%%%%
\begin{figure}[t]
\begin{center}
\includegraphics[width=0.8\linewidth]{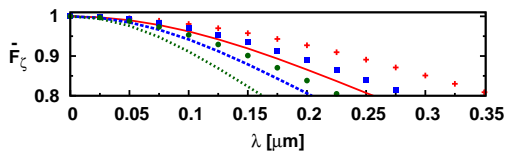}
\caption{(Color online) Average fidelity of harmonic transport versus noise amplitude $\lambda$. For $t_f=20$ ms:  $g_{_{1}}/\hbar =
      0.05\,m/s$ (solid, red line), $g_{_{1}}/\hbar=0.1\,m/s$ (dashed, blue line),
      $g_{_{1}}/\hbar=0.2\,m/s$ (dotted, green line);   
for $t_f=10$ ms: $g_{_{1}}/\hbar =
      0.05\,m/s$ (red crosses), $g_{_{1}}/\hbar=0.1\,m/s$ (blue boxes),
      $g_{_{1}}/\hbar=0.2\,m/s$ (green circles). $\omega_0=2\pi\cdot 50$ Hz.      }
\label{noise}
\end{center}
\end{figure}
%%%%%%%%%%%%%%%%%%%%%%%%%%%%%%%%%%%%%%%%%%%%%%%%%%%%%%%%%%%%%%%%%%%%%  
% 
%
%
In Fig. \ref{fig1} the displacement of the center of mass with respect to the trap center and the trap trajectory
are plotted for this optimal trajectory. 
We have chosen 
$\delta\simeq 0.162$ mm so that the minimal final time is $t_f=20$ ms as in the previous example. 

%Other interesting possibility is to optimize the trajectory $q_0(t)$ 
%to minimize $\int_0^{t_f} |\alpha-q_0| dt$. Proceeding as before 
%the solution for $q_0$ becomes the bang-off-bang direct trajectory
%of the previous section.  

Another important constraint might be that the center of the
physical trap stays inside a given range (e.g. inside the vacuum
chamber), i.e. the constraint is then $q_{\downarrow} \le q_0 (t) \le q_{\uparrow}$.
Following the OCT procedure we finally get
\begin{eqnarray*}
q_0 (t) = \left\{\begin{array}{ll}0,& t=0\\
q_{\uparrow}, &0<t<t_1\\
q_{\downarrow}, &t_1<t<t_f\\
d, &t_f < t
\end{array}\right.,
\end{eqnarray*}
where
$\omega_0 t_1 = \arccos\left[1 - {q_{\downarrow} d - {d^2 \over 2} \over
  q_{\uparrow} (q_{\downarrow} - q_{\uparrow})} \right]$,
%\begin{eqnarray*}
$\omega_0 t_f = \omega_0 t_1 + \arccos\left\{[{q_{\downarrow} d
  - {d^2\over 2} - q_{\downarrow}(q_{\downarrow} - q_{\uparrow})]
  /[(d - q_{\downarrow})(q_{\downarrow} - q_{\uparrow})}] \right\}.$
%\end{eqnarray*}

{\it Noise.}---In the following we investigate the effect of noise in harmonic
transport. We assume that the center of the physical trap is 
randomly perturbed by the shift $\lambda\zeta(t)$ with respect to 
$q_0(t)$.
For the shifted trap center, Eq. (\ref{osc}) can be solved using the ansatz
$\tilde q_c(t) = q_c(t) + \lambda \beta(t)$ so that
$\beta(t)= \int_0^{\omega_0 t} d\tau\, \zeta(\tau) \sin\left(\omega_0 t - \tau\right)$, and 
$\dot\beta(t)= \omega_0 \int_0^{\omega_0 t} d\tau\, \zeta(\tau)
\cos\left(\omega_0 t - \tau\right)$,
with the solution still given by Eq. (\ref{trans}).
The fidelity at $t_F$ is independent of the chosen $q_c$ and $d$, 
${F}_\zeta = \left|\int dq\, \exp\left(\frac{i m}{\hbar} \lambda \dot\beta (t_F)
  q\right) \chi^* (q + \lambda \beta (t_F)) \chi(q)\right|$. %An important
%  property of it is that it is independent of the chosen $\alpha$.
We assume now that $\zeta(t)$ is white Gaussian noise, and average the
fidelity ${F}_\zeta$ over different realizations of $\zeta(t)$.
The result can be seen in Fig. \ref{noise} for three values of $g_{_1}$
and two final times, $t_f=20$ ms and $t_f=10$ ms. 
%For the chosen 
%parameters ${\bar{F}_\zeta}>0.95$ in all three cases for a noise
%amplitude $\lambda < 0.8\, \mu$m. In general, 
The fidelity   
increases for smaller couplings $g_{_{1}}$
and for the shorter time. 

{\it Outlook}---The above results may be extended to other physically motivated constraints, also to non-spherical traps
with different frequencies $\omega_x$, $\omega_y$, $\omega_z$,
rotations, and launching/stopping condensates up to/from a
determined velocity. 
  
%
%{\it{ACKNOWLEDGEMENTS}}
We thank D. Gu\'ery-Odelin for discussions. We acknowledge     
funding by the Basque Government
(Grant No. IT472-10)
and Ministerio de 
Ciencia e Innovaci\'on (FIS2009-12773-C02-01).
E. T. acknowledges financial support from the Basque Government (Grant No. BFI08.151); X. C.  from Juan de la Cierva Programme and the National Natural Science Foundation of China (Grant No. 60806041); S. S. from the German Academic Exchange Service (DAAD).
%We thank A. del Campo, D. Gu\'ery-Odelin, J. Martorell, D. Sprung,
%and E. Sherman for discussions.
%We acknowledge 
%%the kind hospitality of the
%%Max Planck Institute for the Physics of Complex
%%Systems in Dresden, 
%funding by the Basque Government (Grants No. IT472-10 and BFI08.151)
%and Ministerio de 
%Ciencia e Innovaci\'on (FIS2009-12773-C02-01).
%
%
%

\end{document}